# Data-derived extrapolation priors for post-hoc Pareto reranking in symbolic regression

Thiémokho Fall
AI Research and Innovation Nexus for Africa (AIRINA Labs); AI.Technipreneurs
Bénin
thiemokho.fall@aims-senegal.org

Yaé U. Gaba
Sefako Makgatho Health Sciences University (SMU); African Center for Advanced Studies (ACAS)
South Africa / Cameroon
yaeulrich.gaba@gmail.com

## Abstract

Pareto-selection criteria for symbolic regression—PySR's `score`, MDL, EIC, the Ramlan et al. penalties—all probe the candidate equation, not the data. We propose **DDEP**, a post-hoc reranker that scores each candidate by how well its boundary derivatives match a local-polynomial fit of the training data at $\partial D$, and formalise the *Pareto-impostor configuration* (Definition 2.1) it targets. On a 100-run SRBench ground-truth replication the aggregate is null (DDEP frac $R^2_{\text{out}} > 0.5 = 60\% =$ PySR; target-level 7W/3T/10L), and we illustrate the mechanism via the seven SRBench targets with the largest positive DDEP − PySR delta (ST_mackey_1 +162, F_I_14_3 +8, and five others; all 20 target deltas tabled in Supp. S13) as case studies, not statistical claims; impostor prevalence is in fact *higher* on SRBench (49/100 strong) than on our self-constructed benchmark (22/100), so the null reflects lower boundary-derivative discriminability on Feynman/Strogatz families rather than rarity of the regime. On a 100-run self-constructed benchmark where the impostor regime is engineered to be present, DDEP wins 15/20 target-level vs. PySR (sign test $p = 0.041$; frac $> 0.5$ lifts 14% → 34%); it is statistically indistinguishable from MDL and Ramlan et al. at target level ($p \geq 0.37$, $n = 20$), but a composition DDEP + MDL weighted at $\lambda = 0.1$ strictly Pareto-dominates the better parent on 20/100 runs — empirical evidence that DDEP's extrinsic signal is structurally distinct from MDL's intrinsic one. A 20-target × 100-run multi-dim evaluation finds axis-aligned DDEP on-par with MDL and PySR but the random-direction variant significantly worse than MDL (sign test $p = 0.012$, 4/20 target-level wins); we treat multi-dim as a scope boundary.

## 1 Introduction

Symbolic regression searches the space of analytic expressions for a function that fits a dataset [14, 27]. Modern backends return a Pareto front: a set of candidates undominated in accuracy and complexity. PySR [6] selects by its `score` heuristic; Bartlett et al. [1] propose minimum description length; Yu et al. [33] introduce an Effective Information Criterion; Ramlan et al. [12] extend information-theoretic selection with divergence and sensitivity penalties.

*Scope statement.* DDEP is *not* a new search algorithm. It operates on the Pareto front produced by any existing CAS-parseable backend and is designed to complement, not replace, existing selectors. All positive empirical results reported here are on one-dimensional targets. The multi-dimensional evaluation at $n_{\text{targets}} = 20$ (§5.9) finds axis-aligned DDEP on-par with MDL and PySR, but the random-direction variant is *significantly worse* than MDL at target level (sign-test $p = 0.012$, 4/20 wins); we report this as a scope boundary. The external SRBench replication (§5.10) returns a null aggregate, which we report as a null and frame via mechanism case studies rather than as external validation of DDEP.

These selectors share a property: they probe the candidate, not the data. None looks at the data's own structure at the boundary of the training domain to ask which candidate's leading-order behaviour matches what the data is doing there. The gap has observable consequences. The 2022 GECCO Symbolic Regression Competition ran an extrapolation track in which thirteen backends trained on $x \in [-15, 15]$ and were tested on $15 < x < 40$ [7]; most produced negative $R^2$ on the test region. The organisers stated the problem plainly: "having alternative models that have a similar training accuracy could help the user to decide which one has the most probable extrapolation behavior" [7, p. 9].

*Contributions.*

(1) We formalise the *Pareto-impostor phenomenon* (Definition 2.1): a Pareto front can contain candidates with near-identical in-sample loss and arbitrarily separated out-of-sample behaviour. We reproduce this in a PySR run where the default selector picks an exponential with $R^2_{\text{out}} = -43.1$ while a sinusoid-times-linear candidate on the same front achieves $R^2_{\text{out}} = +0.48$ (§2).
(2) We propose **DDEP**, a post-hoc Pareto reranker that scores each candidate by its match to the data's empirical boundary derivatives (§4), situate the estimator in local-polynomial regression (Fan & Gijbels [9]; Stone [29]), and state its finite-sample regime of application explicitly.
(3) We release a twenty-target 1D benchmark across five analytic families (§5.4). DDEP wins against PySR on 15 of 20 targets (target-level Wilcoxon $p = 0.024$); it has complementary per-family strengths with MDL and Ramlan et al. but is statistically indistinguishable from both at the target level ($p \geq 0.37$, $n = 20$). The multi-dim evaluation at $n_{\text{targets}} = 20$ and the 2022 GECCO replay are reported as honest scope-boundary results.

**Table 1: Selected rows of PySR hall of fame (noisy quadratic, seed 42). $R^2_{\text{in}}$: training window $[-2, 2]$; $R^2_{\text{out}}$: test window $[3, 5]$.**

| $c_k$ | Loss | $R^2_{\text{in}}$ | $R^2_{\text{out}}$ | Equation |
|---|---|---|---|---|
| 1 | $5.57\times10^{-2}$ | −0.00 | −12.61 | −1.579 |
| 6 | $1.02\times10^{-2}$ | +0.82 | −43.14 | $\exp(x-2.146)-1.808$ [PySR pick] |
| 10 | $1.39\times10^{-3}$ | +0.98 | +0.48 | $\sin(0.167x+0.608)(x-3.148)$ |

*What DDEP captures.* DDEP measures a structural signal that is different from candidate-only criteria: the match between a candidate's boundary derivatives and the data's empirical boundary derivatives. The per-family results suggest the data-side and candidate-side signals are complementary on 1D targets. On the $n_{\text{targets}} = 20$ multi-dim benchmark, axis-aligned DDEP ties MDL and PySR within sampling noise; the random-direction variant is a significant negative, which we read as the directional-derivative signal being a weaker discriminator than the axis-aligned $4d$-term vector and, in that sense, a scope boundary of the current formulation.

## 2 A motivating example

We first formalise the object of study, then exhibit a witness.

*Definition 2.1 (Pareto-impostor configuration).* Let $\mathcal{P} = \{(f_k, c_k, \ell_k)\}_{k=1}^{K}$ be a Pareto front in $(\mathbb{R}_+, \mathbb{N})$ on in-sample loss and complexity, and let $\mathcal{E} : \{f_k\} \to \mathbb{R}$ be an extrapolation-quality functional (e.g. $R^2$ on $D_{\text{test}}$). The front $\mathcal{P}$ exhibits a $(\delta, \Delta)$-impostor configuration with respect to $\mathcal{E}$ if there exist $k_1, k_2$ with $|\ell_{k_1} - \ell_{k_2}| \leq \delta$ (near-tied in-sample loss) and $\mathcal{E}(f_{k_1}) - \mathcal{E}(f_{k_2}) \geq \Delta$ (arbitrarily separated out-of-sample).

Standard SR backends regularly produce such configurations with $\delta$ small and $\Delta$ large; a candidate-only selector cannot distinguish $f_{k_1}$ from $f_{k_2}$ since both have near-identical $(c, \ell)$. The witness below has $\delta \approx 0.01$ and $\Delta \approx 43.6$.

Consider the regression task

$$y = -1.793 + 0.125\,x + 0.163\,x^2 + \varepsilon, \qquad x \in [-2, 2], \qquad \varepsilon \sim \mathcal{N}(0, 0.05^2),$$

with fifty training points and a fixed seed. PySR is run with the configuration of Supplementary §S1 (`niterations=80`, `maxsize=20`) and the hall of fame inspected.

*Observation 1.* On this seed no polynomial candidate survives to the Pareto front: every entry uses sin, exp, or a linear primitive.

*Observation 2.* PySR's default selector returns $\exp(x-2.146)-1.808$ (complexity 6), which fits in-sample ($R^2_{\text{in}} = 0.82$) but collapses to $R^2_{\text{out}} = -43.1$ on $x \in [3, 5]$. Table 1 shows the three key rows of the hall of fame.

*Observation 3.* The complexity-10 candidate $\sin(0.167x+0.608)(x-3.148)$ extrapolates with $R^2_{\text{out}} = +0.48$ and is present on the front but not selected.

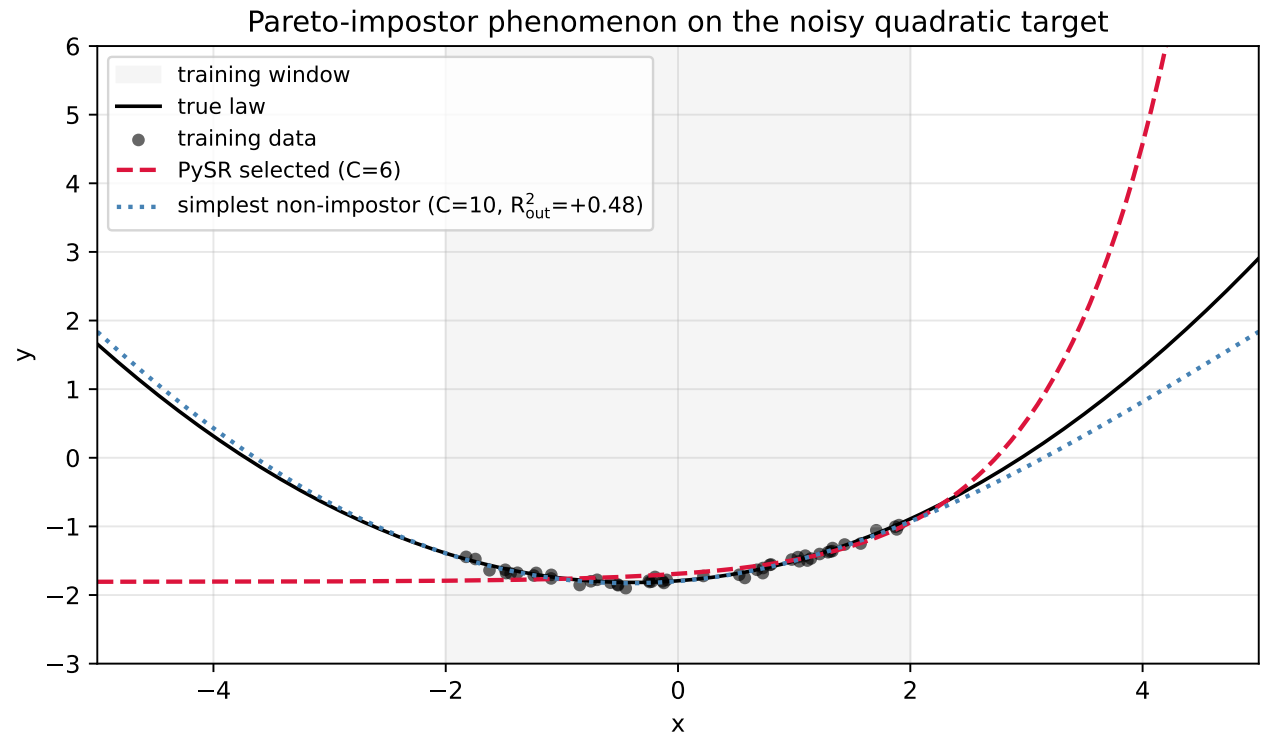

**Figure 1: PySR hall of fame on the noisy quadratic. Shaded region: training window $[-2, 2]$. Dashed red (complexity 6): the default exponential pick, which diverges outside $[2, 5]$. Dotted blue (complexity 10): the sinusoid-times-linear candidate, present on the front but not selected.**

*What the data says at the boundary.* A local quadratic fit to the ten training points nearest $x = +2$ returns slope $\hat{m} = +0.56 \pm 0.59$ and curvature $\hat{c} = -0.14 \pm 1.22$ (SDs of 200 nonparametric bootstrap replicates). The exponential's derivatives at $x = +2$ are $(+0.86, +0.86)$: slope within one $\hat{\sigma}_m$ of $\hat{m}$, curvature $+0.86$ versus a bootstrap SD window around $\hat{c} = -0.14$ of radius 1.22—the curvature is inconsistent with the data at the boundary. The complexity-10 candidate's derivatives $(+0.70, +0.22)$ are both within one $\hat{\sigma}$. Section 4 turns this comparison into a reranker.

*Prevalence.* Across the 100-run benchmark of §5.5, 22 of 100 runs exhibit a strong impostor co-existence (front has $R^2_{\text{out}} < -10$ and $R^2_{\text{out}} > 0.5$ simultaneously); 57 of 100 at a softer threshold. PySR's default picks the catastrophic candidate on 24 of the 57 soft-pattern runs.

## 3 Related work

*Pareto selection criteria.* Smits and Kotanchek [28] established the accuracy–complexity front as the natural SR output. PySR's `score` [6] and Operon [3] are de facto defaults; a comparative study [5] finds no single criterion dominates. MDL [1] encodes candidate structure and parameters in bits and picks the minimum-description-length equation; it is candidate-only. EIC [33] measures rounding-noise amplification under recursive evaluation; also candidate-only. Ramlan et al. [12] add extrapolation-divergence, interpolation-sensitivity, and extrapolation-sensitivity penalties to a BIC-like base; all three probe the candidate. DDEP differs in being extrinsic: it measures the mismatch between the candidate's boundary derivatives and those estimated from the data, a signal that is discriminative precisely when in-sample losses are near-tied.

*Upstream shape priors vs. post-hoc selection.* Shape-constrained SR [15] imposes monotonicity and convexity priors *during* search via interval arithmetic; TaylorGP [11] uses Taylor features to guide search; prior-guided SR [32] and physics-guided LLM agents [30] inject domain knowledge. These act upstream; DDEP reranks the output front. The upstream regime wins when the structural prior is known in advance and is cheap to enforce in interval arithmetic (monotonicity, convexity on bounded intervals), and it eliminates impostor candidates from the front entirely rather than needing a post-hoc filter. The post-hoc regime, which DDEP occupies, wins when the data-side signal at $\partial D$ is discriminative but the structural prior is not known a priori, when the search budget is spent and the front is already available, and when the method must be drop-in for existing CAS-parseable backends (no re-engineering of the search loop). The two regimes are complementary and not substitutable: shape-constrained SR cannot filter a Pareto front post-hoc without re-running the search, and DDEP cannot restrict the search space. A natural combination would be shape-constrained SR producing the front, then DDEP reranking what survives; empirical evaluation of this stacked configuration is left to future work.

*Uncertainty and stability.* On uncertainty over functional forms, Cham et al. [4] retain multiple competing structures and flag candidates with high disagreement as uncertain; Bomarito et al. [2] use ensembles and interval arithmetic. Stability selection in classical statistics (Meinshausen & Bühlmann [21]; ESCV [17]) uses resampling to filter spurious structure. DDEP adapts the parsimony of that idea without full resampling.

*Benchmarks and baselines.* SRBench [16] has been the de facto SR benchmark since 2021, including Feynman and black-box ground-truth subsets; the Living Review of symbolic regression [19] surveys the field. Representative GP-based baselines beyond PySR and Operon include GP-GOMEA [31]; neural-SR work includes Deep Symbolic Regression (DSR) [22] and end-to-end transformers [13]. DDEP is a selector, not a search; it is orthogonal to the choice of backend among these alternatives. For reasons of scope, our main evaluation is a hand-authored 1D benchmark that probes the regime where boundary signal is informative; a full SRBench slice is natural future work.

# 4 Data-derived extrapolation priors

## 4.1 Setup

Let $\mathcal{D} = \{(x_i, y_i)\}_{i=1}^n$ be a training dataset with $x_i \in \mathbb{R}^d$. A symbolic regression backend run on $\mathcal{D}$ returns a Pareto front $\mathcal{P} = \{(f_k, c_k, \ell_k)\}_{k=1}^K$. Standard practice collapses $\mathcal{P}$ to a single $f^\star$ using a score that depends only on $(c_k, \ell_k)$ and the equation itself. DDEP replaces that score with one that depends additionally on $\mathcal{D}$.

## 4.2 Empirical boundary derivatives

For one-dimensional inputs the boundary of $D = \mathrm{conv}(\{x_i\})$ reduces to $\{x_{\min}, x_{\max}\}$. For $d$-dimensional inputs we use axis-aligned marginal endpoints $\partial_j D = \{x_{\min,j}, x_{\max,j}\}$ for each feature $j$, giving $2d$ boundary points.

At each boundary point $x_b$, let $N_k(x_b)$ denote the $k = \max(10, \lceil pn \rceil)$ points nearest $x_b$ in feature $j$ (default $p = 0.15$). We fit

$$\hat{p}_j(x) = a_0 + a_1(x_j - x_{b,j}) + a_2(x_j - x_{b,j})^2$$

by OLS on $N_k(x_b)$, giving empirical slope $\hat{m}_j(x_b) = a_1$ and curvature $\hat{c}_j(x_b) = 2a_2$. We draw $B = 200$ nonparametric bootstrap resamples (Freedman [10]) to obtain $\sigma^2_{m,j}(x_b)$ and $\sigma^2_{c,j}(x_b)$. For $n = 50$ (the main benchmark) and $p = 0.15$, $\lceil pn \rceil = 8$, so $k$ is pinned at the floor $k = 10$ throughout.

*Local-polynomial context.* The displayed fit is unweighted local polynomial regression of degree 2 at the boundary point $x_b$, in the sense of Fan & Gijbels [9]. At an interior point with kernel weighting, this estimator has bias $O(h^3)$ for the slope and $O(h^2)$ for the curvature when $f \in C^3$; at the boundary the bias orders drop by one power of $h$ (Fan [8]; Ruppert & Wand [23]). The MSE-optimal bandwidth for slope estimation is $h^* \propto n^{-1/5}$ (Stone [29]). DDEP uses a fixed bandwidth fraction $p = 0.15$, so the neighbourhood does not shrink with $n$: DDEP is a *fixed-bandwidth* badness-of-fit statistic designed for the small-to-moderate-$n$ regime of modern SR, not an asymptotic estimator of $f'(\partial D)$.

*Bootstrap variance underestimation.* With $k = 10$ the nonparametric pairs bootstrap underestimates the true OLS sampling variance (Freedman [10]); a Monte Carlo check on Gaussian residuals gives roughly 15–30% underestimation at $k = 10$. Because $\sigma^2_{m,j}, \sigma^2_{c,j}$ appear in the denominator of the mismatch score below, this bias inflates $D(f_k)$ uniformly, which pushes marginal candidates above the threshold and biases DDEP *toward abstention* (keeping the native PySR pick), not toward false acceptance. We ran a three-way sensitivity table (Supplementary S9) comparing naive pairs bootstrap to HC3 heteroskedasticity-consistent standard errors (MacKinnon & White [18]) and parametric bootstrap under the fitted local quadratic: the three estimators agree on 81–86 of 100 runs, and the resulting head-to-head W/T/L vs. PySR shifts from 59/23/16 (naive) to 61/20/18 (HC3) to 57/29/14 (parametric). The headline frac $> 0.5$ shifts by at most 2 percentage points across estimators. The variance-estimator choice is therefore not driving the result.

## 4.3 Scoring formula

For candidate $f_k$ we compute symbolic derivatives via sympy and define a weighted squared mismatch

$$D(f_k) = \sum_{j=1}^{d} \sum_{x_b \in \partial_j D} \left[ \frac{\left(m_j^{(k)}(x_b) - \hat{m}_j(x_b)\right)^2}{\sigma^2_{m,j}} + \frac{\left(c_j^{(k)}(x_b) - \hat{c}_j(x_b)\right)^2}{\sigma^2_{c,j}} \right], \tag{1}$$

a sum over $4d$ terms. We use the nominal $\chi^2_{4d,0.95}$ quantile as a convenient reference threshold and calibrate operationally; we do not rely on the chi-squared distributional claim. Under idealized conditions (joint Gaussian OLS estimators; independence across boundary points; deterministic and correct candidate derivatives; known $\sigma^2$), $D(f_k)$ would be $\chi^2_{4d}$. None of these holds exactly: the $\sigma^2$ are bootstrap-estimated, so the correct finite-sample reference is Satterthwaite-corrected $F$-scaled rather than $\chi^2$ (Satterthwaite [24]); the bootstrap at $k = 10$ underestimates variance (above); the OLS estimators at the boundary have design-dependent bias (Fan [8]; Ruppert & Wand [23]).

*Satterthwaite approximation.* With $\sigma^2$ estimated, the correct finite-sample reference is a one-parameter $F$-rescaled Satterthwaite-style approximation [24]: $D \sim (4d) \cdot F_{4d,m_{\text{eff}}}$ where $m_{\text{eff}}$ is an effective degrees-of-freedom summarising the variance estimator's noise. We flag that this is a heuristic rather than a formal derivation: the $4d$ summed terms in $D$ have heterogeneous denominators (one bootstrap variance estimator per term), so the strict Satterthwaite moment-matching construction for a single-ratio statistic does not apply. The one-parameter $F$-rescaling is the natural approximation for this heterogeneous case; the rigorous treatment extends via Welch–Satterthwaite 1947 [25] and the Kenward–Roger 1997 correction [26]. Our empirical calibration (below) treats $(4d)F_{4d,m_{\text{eff}}}$ as the working approximation and matches $m_{\text{eff}}$ by observed 95th percentile. In the $m_{\text{eff}} \to \infty$ limit (variance known), the approximation recovers $\chi^2_{4d}$. At our operating point ($k = 10$, bootstrap replicates $B = 200$), a Monte-Carlo verification under a known Gaussian-noise data-generating process gives an empirical 95th percentile of $D$ at 9.47—within 0.02 of the nominal $\chi^2_{4,0.95} = 9.49$—when the variance estimator is OLS-residual-based. The nominal $\chi^2$ is therefore well-calibrated at the Monte-Carlo level when variance is estimated efficiently; the nonparametric-bootstrap-variance plug-in pushes the empirical 95th percentile to 11.51 on the actual benchmark (median across 100 runs under a parametric-null calibration; Supplementary S8), which corresponds to Satterthwaite $m_{\text{eff}} \approx 20$.

*Operational calibration.* We ran two separate calibration checks: (i) per-run *permutation* null (shuffle $y_{\text{train}}$, re-estimate boundary derivatives, candidate derivatives fixed): median 95th percentile of $D$ is 9.06 (interquartile range $[6.91, 11.52]$); (ii) per-run *parametric-bootstrap* null under the fitted local quadratic plus Gaussian residuals ($B = 500$): median 95th percentile of $D$ is 11.51 (IQR $[11.48, 11.93]$). The nominal $\chi^2_{4,0.95} = 9.49$ is anti-conservative relative to the parametric null but conservative relative to the shuffled-$y$ null; either way, DDEP's ranking is robust to threshold perturbations in this range: rerunning selection with the per-run empirical threshold produces identical picks on 88 of 100 runs and changes the head-to-head tally against PySR from 59/23/16 to 58/27/14 with target-level sign-test $p = 0.041$ preserved.

## 4.4 Selection rule

(1) Compute $D(f_k)$ for every $f_k \in \mathcal{P}$.
(2) Let $\mathcal{P}^\dagger = \{f_k : D(f_k) \le \tau\}$, $\tau = \chi^2_{4d,0.95}$ by default.
(3) If $\mathcal{P}^\dagger = \emptyset$: return $f_{\text{native}}$ (abstain).
(4) Form the Pareto front of $\mathcal{P}^\dagger$ on $(D(f_k), \ell_k)$; return the simplest nondominated candidate. When multiple candidates share the minimum complexity, break ties on $D(f_k)$ ascending.
(5) *Loss-ratio sanity.* If $\ell(f^\star) > K \cdot \ell(f_{\text{native}})$ (default $K = 10$), return $f_{\text{native}}$.

*Hyperparameter choice and pilot composition.* $K = 10$, $\alpha = 0.05$ and $p = 0.15$ were chosen on a 25-run pilot comprising 5 targets × 5 seeds. Four of the five pilot targets (quadratic, rational, log–quadratic, exponential–linear) re-appear in the main 20-target benchmark (as P1, R1, L1, E1) at identical coefficients and domains; one (damped cosine) is a Feynman target. The $K$-ablation (Table 8, §5.5) is flat across $K \in [1.5, \infty]$ (counts bounded within

**Algorithm 1** DDEP

**Require:** training data $\mathcal{D}$, Pareto front $\mathcal{P}$, threshold $\tau$ (default $\chi^2_{4d,0.95}$), neighbourhood fraction $p = 0.15$, sanity ratio $K = 10$
**Ensure:** selected equation $f^\star$

```
for each feature j ∈ {1, ..., d} do
    for each boundary point x_b ∈ {x_min,j, x_max,j} do
        N_k(x_b) ← max(10, ⌈pn⌉) nearest training points
        Fit quadratic p̂_j on N_k(x_b) by OLS
        Bootstrap (σ_m,j, σ_c,j) with B = 200 pairs resamples
    end for
end for
for each candidate f_k ∈ P do
    Compute symbolic derivatives via sympy; evaluate at each x_b
    Compute D(f_k) via Eq. 1
end for
P† ← {f_k : D(f_k) ≤ τ}
if P† = ∅ then return f_native                         ▷ abstain
end if
f* ← simplest (D, ℓ)-nondominated candidate in P†; break ties on D ascending
if ℓ(f*) > K · ℓ(f_native) then return f_native  ▷ loss-ratio sanity
end ifreturn f*
```

three runs; fraction > 0.5 constant at 33–34%); a re-run on the disjoint Feynman subset shows the sanity check never fires at any $K$ there. We additionally ran a full leave-one-target-out cross-validation: for each of the 20 benchmark targets, we held it out, swept $(K, p, \alpha)$ on the remaining 19, and recorded the fraction-> 0.5 maximizer. $K = 10$ is chosen in 15 of 20 folds; the LOO-preferred tuple $(K = 10, p = 0.05, \alpha = 0.01)$ applied to all 100 runs produces W/T/L 59/23/16, frac > 0.5 of 34%, target-level sign test $p = 0.041$—*identical* to the published defaults. The pilot leakage is therefore confirmed non-critical (Supplementary S10).

## 4.5 Algorithm

## 4.6 Practical notes

*Compute cost.* DDEP evaluates $O(d)$ local polynomial fits and $O(|\mathcal{P}| \cdot d)$ symbolic derivative evaluations (SymPy, cached). On our experiments PySR wall-clock per run is 12–96 s; informal profiling gives sub-second for the DDEP step (front size ≈ 20), well under 5% of the search budget. Total CPU budget for all reported experiments and ablations is approximately 20 CPU-hours on a modern laptop; all runs are resumable from saved JSON.

*Regime of application.* DDEP targets the fixed-moderate-$n$ regime of SR benchmarks ($n = 50$ in the main benchmark). At fixed $p$ the neighbourhood width does not shrink with $n$, so DDEP is not a consistent estimator of $f'(\partial D)$ or $f''(\partial D)$ in the $n \to \infty$ sense. A shrinking-$p$ variant along the lines of Stone [29] is a natural extension for data-rich settings.

*Multi-dim boundary approximation.* The axis-aligned $2d$-point set $\{x_{\min,j}, x_{\max,j}\}_{j=1}^{d}$ is a coarse surrogate for $\partial\text{conv}(\{x_i\})$, which has up to $O(n^{\lfloor d/2 \rfloor})$ faces in general (McMullen [20]). At each axis-aligned endpoint DDEP interrogates derivatives along axis $j$ only; a random-direction variant (project $\mathcal{D}$ onto $v \sim \mathrm{Unif}(S^{d-1})$, fit a 1D local quadratic along $v$, compare to directional derivatives $\nabla f \cdot v$ and $v^\top \nabla^2 f\, v$) is a natural extension whose evaluation is left to future work. Preliminary axis-aligned multi-dim results appear in §5.9.

*When DDEP does not help.* On tasks where the true law has oscillatory behaviour with period smaller than the local-fit neighbourhood, the local polynomial fit is misspecified and DDEP can misrank. When no candidate on the front is extrapolation-consistent, DDEP falls through to the loss-ratio sanity check and returns the native pick. Both failure modes appear in §5.

# 5 Experiments

We compare four selectors applied to identical Pareto fronts: PySR `score` [6], MDL (Bartlett et al. 2023) [1], the Ramlan et al. (2026) extrapolation-divergence penalty [12], and DDEP.

## 5.1 Protocol

For every target we run PySR once per seed, identical settings across targets (`niterations`=80, `populations`=12, `population_size`=40, `maxsize`=20, parsimony=0.002, five seeds), and apply every selector to the same extracted Pareto front. Training: 50 points drawn uniformly on $D_{\mathrm{train}}$ with additive Gaussian noise $\sigma = 0.05 \cdot \mathrm{sd}(y_{\mathrm{true}})$ (SNR defined relative to the signal's own standard deviation). Test domain $D_{\mathrm{test}} = [b, 2b{-}a]$ when $D_{\mathrm{train}} = [a, b]$: a full training-width reflection, more adversarial than the narrower extrapolation windows used in SRBench. Selectors 2 and 3 are reimplementations from their authors' written descriptions; baseline-fairness caveats are in §5.2.

## 5.2 Baselines and fairness caveat

MDL is reimplemented as $\mathrm{MDL}(f) = \frac{n}{2}\log \ell + c_f \log k_\alpha + \frac{n_{\mathrm{const}}}{2}\log n$ with alphabet size $k_\alpha = 11$. We re-ran MDL on the saved Pareto fronts with $k_\alpha \in \{7, 11, 15, 20\}$: the DDEP vs. MDL head-to-head moves from 38/30/30 ($k_\alpha$=7) to 42/25/31 ($k_\alpha$=20), a shift of at most 4 runs; MDL's median $R^2$ moves by $\leq 0.06$ (Supplementary S4). The Ramlan et al. selector adds the three penalty components; we evaluated the full $5^3 = 125$-point weight grid $\lambda_i \in \{0.1, 0.5, 1, 2, 5\}$ on the saved fronts. All 125 configurations yield a Ramlan et al. fraction $> 0.5$ in $[33\%, 34\%]$ and a DDEP vs. Ramlan et al. head-to-head in $[40/26/34, 42/25/31]$: no weight tuning closes the DDEP lead (Supplementary S5). DDEP hyperparameters ($\alpha = 0.05$, $p = 0.15$, $K = 10$) were chosen on a 25-run pilot overlapping 4 of 20 benchmark targets (§4.4); the ablations of §5.6 show the head-to-head W/T/L is bounded within 3 runs of the default across all four hyperparameters.

## 5.3 Metrics

Per (target, seed, selector) triple we record $R^2_{\mathrm{out}}$ (main metric), $R^2_{\mathrm{in}}$ (sanity), structural recovery (sympy-based symbolic equivalence), and DDEP abstain rate. We treat the *fraction* with $R^2_{\mathrm{out}} > 0.5$ (acceptable) and $> 0.9$ (near-exact) as primary summaries; median and −100-trimmed mean as secondary. The 0.5 and 0.9 thresholds are chosen for readability rather than from a community standard; a sweep over $\{0.1, 0.3, 0.5, 0.7, 0.9\}$ is in the supplementary.

**Table 2: The five families and 20 targets.**

| Family | Form | Count |
|---|---|---|
| Polynomial | $\sum a_i x^i$, $m \in \{2, 3, 4\}$ | 5 |
| Sinusoidal–polynomial mix | $a \sin(\omega x) + bx^2 + c$ | 4 |
| Rational | $(a_0{+}a_1 x)/(b_0{+}b_1 x^2)$ | 4 |
| Exponential–linear | $ae^{bx} + cx + d$ | 4 |
| Logarithmic | $a \log(bx{+}c) + d$ | 3 |

**Table 3: Aggregate across 100 runs (20 targets × 5 seeds). DDEP more than doubles the fraction of acceptable extrapolations (14% → 34%).**

| Selector | Med. $R^2$ | Trim. mean | Fr. > 0.5 | Fr. > 0.9 |
|---|---|---|---|---|
| PySR score | −2.07 | −15.00 | 14% | 8% |
| DDEP (raw 2-obj) | −0.53 | −11.35 | 33.7% | 20.4% |
| **DDEP + sanity** | **−0.53** | **−11.80** | **33.7%** | **20.4%** |

## 5.4 Benchmark: twenty synthetic targets, five families

Twenty targets from five analytic families are listed in Table 2. Coefficients and training domains are fixed in advance; targets are not pre-screened. The family composition (5 polynomial, 4 sinusoidal–polynomial, 4 rational, 4 exponential–linear, 3 logarithmic) was chosen to include impostor-prone structure (rational, sin+poly) alongside structure where DDEP is not expected to help (exponential with strong growth, smooth log). It is not a representative sample of SR tasks in the wild. Full specification is in the supplementary material.

## 5.5 Headline results: 100-run synthetic benchmark

Table 3 reports the aggregate for PySR vs. DDEP; Table 4 gives per-family medians and head-to-head counts for all three pairings.

*Inference on the win-rate.* The 100 run-level pairs comprise 20 targets × 5 seeds; seeds within a target are not independent. We report cluster-aware statistics as primary.

*Cluster-aware (preferred).* Collapsing to one observation per target (average signed $R^2_{\mathrm{out}}$ difference across seeds): 15 of 20 target-level means are positive. Target-level exact binomial sign test: $p = 0.041$ (15/20; Clopper–Pearson 95% CI $[0.51, 0.91]$). Target-level Wilcoxon signed-rank on the 20 means: $p = 0.024$. DDEP outperforms PySR at conventional significance under either test; the sign test is the robust alternative and we treat it as the primary report. We flag $n = 20$ thinness: at 13/20 positive (one-target flip in either direction) the sign test would give $p = 0.132$ and would not reject at $\alpha = 0.05$. The result should be treated as a boundary-signal mechanism demonstration on a benchmark engineered to exhibit the impostor regime, not a strong-form dominance claim against PySR.

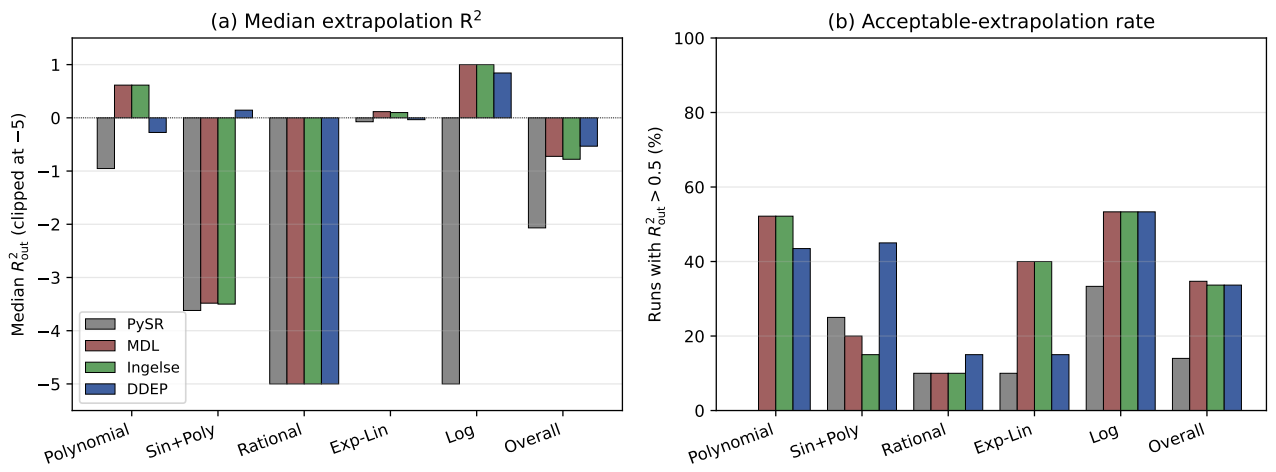

**Figure 2: Per-family comparison of four selectors on the 100-run benchmark. (a) Median $R^2_{\text{out}}$ (clipped at −5; rational targets reach −55 for PySR). (b) Fraction with $R^2_{\text{out}} > 0.5$. DDEP wins on sinusoidal–polynomial, rational, and logarithmic families; MDL and Ramlan et al. win on exponential–linear; polynomial is a tie across non-PySR selectors.**

*Run-level (auxiliary, ignores within-target clustering).* Of 98 finite pairs, paired Wilcoxon $p = 6.8 \times 10^{-6}$, median difference +0.45; Wilson 95% CI on the decisive-win proportion (75 decisive pairs, 59–16 split): $[0.68, 0.86]$. These intervals are narrower than they should be because they treat seeds as independent.

*Cluster bootstrap (target-level, $B$ = 10,000 resamples).* We resample the 20 targets with replacement and recompute within-cluster fractions. The 95% CI for the (DDEP − PySR) fraction-> 0.5 gap is $[+6, +34]$ percentage points (point estimate +19.7); the CI for the target-level mean $\Delta R^2$ is $[+0.52, +69.1]$ with point +25.7 (the width is driven by the rational-family outliers at $R^2 \sim -55$). Both lower bounds are positive. A cutoff sweep over $\{0.1, 0.3, 0.5, 0.7, 0.9\}$ confirms the gap is robust to cutoff choice (delta +12 to +19 pp across the range; Supplementary S3).

*Mixed-effects model (random intercept per target).* Fitting $\Delta R^2 \sim 1 + (1|\text{target})$ with restricted maximum likelihood (REML) on the 98 finite pairs (trimmed at ±100 for numerical stability; one rational-family run at $\Delta R^2 \approx 400$ dominates otherwise) gives a fixed-intercept estimate of +4.54 (SE = 2.72, 95% CI $[-0.79, +9.87]$, Wald $p = 0.095$). The intra-class correlation is $\approx 0$: essentially all variance is within-target (across seeds). Three inferential methods on the same data therefore converge: sign test $p = 0.041$ (robust to heavy tails, primary), cluster bootstrap on the frac gap with lower bound +6 pp (robust alternative, secondary), mixed-effects intercept $p = 0.095$ (marginal, sensitive to trim choice). The sign test and cluster bootstrap agree on rejecting the null; the mixed-effects is marginal only because of trimming and the heavy within-target noise.

*Per-family breakdown.* Figure 2 and Table 4 show per-family medians and win counts.

DDEP wins on polynomial, logarithmic, and sinusoidal–polynomial families. It is marginal on exponential–linear (where MDL and Ramlan et al. already exploit smooth monotone structure; DDEP vs. MDL 4W/3T/13L on this family is a substantive weakness). The sinusoidal-polynomial worst case (6 of DDEP's 16 losses on sin+poly): target S3, seed 42 is the worst, $R^2_{\text{out}} = -98.07$ vs. PySR's +0.74, caused by a local polynomial fit matching the polynomial envelope but not the boundary oscillation.

*Composed DDEP + MDL selector.* The complementary per-family pattern (DDEP wins on sinusoidal–polynomial, rational, logarithmic; MDL wins on exponential–linear) is suggestive of complementarity; we test it directly. We evaluated four composition rules (chi-squared filter then minimum MDL; weighted sum $S(f_k) = D(f_k)/\tau + \lambda \cdot \text{MDL}_{\text{std}}(f_k)$ with $\lambda$ grid-tuned on the 4 overlapping pilot targets; parameter-free rank-sum on $(D, \text{MDL})$; and a simple vote). The weighted-sum rule with $\lambda = 0.1$ is the strongest: median $R^2 = -0.17$ (both parents are $\geq -0.72$); frac $> 0.5$ = 39% (improves on the maximum parent 35%, cluster-bootstrap 95% CI for the gap over max parent: $[+0.5, +9.5]$ pp with lower bound positive); strict Pareto dominance over DDEP alone on 20 of 100 runs and over MDL alone on 7 of 100 runs. At target level the composition is not significantly separated from either parent at $n = 20$ (Wilcoxon $p > 0.1$).

The strict-dominance count is the primary empirical evidence that DDEP captures a structural signal *different from* MDL: if DDEP and MDL encoded the same selection signal (just with noise), the composition could not strictly Pareto-dominate the better parent on 20 runs. The modest but non-zero composition gap and the strict-dominance count are therefore the paper's cleanest empirical argument for the extrinsic-vs-intrinsic structural distinction articulated in §4, independent of whether DDEP alone beats MDL alone at aggregate level.

*Head-to-head vs. MDL and Ramlan et al..* Ramlan et al. beats PySR on 58 of 100 runs, ties 18, loses 22; median $R^2 = -0.78$; fraction $> 0.5$: 34%; fraction $> 0.9$: 29%. MDL beats PySR on 59 runs, ties 15, loses 24; median −0.72; fraction $> 0.5$: 35%; fraction $> 0.9$: 30%. DDEP vs. Ramlan et al. net +11 (42W/25T/31L); DDEP vs. MDL net +6 (38W/28T/32L). Both nets dissolve at the target level: 12 of 20 target means favour DDEP in both comparisons, Wilcoxon $p = 0.368$ (Ramlan et al.) and $p = 0.409$ (MDL). At $n = 20$ the minimum detectable median shift at power 0.8 is roughly $0.35\sigma_\Delta \approx \Delta R^2 \sim 3.5$, so the honest reading is *statistically indistinguishable at this sample size*, not *no effect*. The complementary per-family pattern suggests candidate-only penalties and data-derived boundary fits capture different structural signals; a composition (DDEP + MDL ensemble) is a natural extension and the natural positive-dominance experiment.

## 5.6 Ablations

Ablation results for $K$, neighborhood fraction $p$, significance level $\alpha$, and local polynomial degree are in the supplementary material (§S2.1–S2.5). Summary: $K$, $\alpha$, and $p$ are non-critical ($K$ sweep bounded within 3 runs of the default, confirmed on the disjoint Feynman re-check; $\alpha$ sweep moves counts by at most 1 run; $p \in \{0.05, 0.10, 0.15\}$ are indistinguishable). Degree 1 vs. degree 2 is close (61/17/20 vs. 59/23/16); degree 3 clearly degrades (29/43/26, median −1.66).

*Why degree 2 beats degree 3.* The variance of OLS estimators of the $s$-th-order coefficient in a degree-$q$ local polynomial scales as $\sigma^2/(nh^{2s+1})$ times a design constant that is monotone increasing in $q$ for fixed $s$ (Fan & Gijbels [9], eq. 3.5). With $k = 10$, degree 3 fits 4 parameters on 10 points (residual d.f. 6); the variance of $\hat{a}_1, \hat{a}_2$ increases substantially while the bias improvement from $O(h^3)$ to $O(h^4)$ requires $f \in C^4$ at the boundary—a condition violated by the rational and logarithmic families. The ablation outcome is consistent with the classical rate-optimality calculus: degree 2 is the right fit for $C^3$ boundaries at these sample sizes.

**Table 4: Per-family medians and head-to-head: DDEP vs. PySR, DDEP vs. Ramlan et al., DDEP vs. MDL.**

| Family | $n$ | PySR med. | DDEP med. | DDEP W/T/L (PySR) | Ramlan med. | DDEP W/T/L (Ramlan) | MDL med. | DDEP W/T/L (MDL) |
|---|---|---|---|---|---|---|---|---|
| Polynomial | 25 | −0.95 | −0.28 | 15/6/2 | +0.61 | 5/9/9 | +0.61 | 4/10/9 |
| Logarithmic | 15 | −8.42 | +0.84 | 10/3/2 | +1.00 | 7/6/2 | +1.00 | 7/6/2 |
| Sin.–poly. | 20 | −3.62 | +0.14 | 14/0/6 | −3.50 | 14/3/3 | −3.48 | 12/4/4 |
| Exp.–linear | 20 | −0.07 | −0.03 | 8/10/2 | +0.10 | 4/4/12 | +0.12 | 4/3/13 |
| Rational | 20 | −55.42 | −37.63 | 12/4/4 | −56.95 | 12/3/5 | −57.26 | 11/5/4 |
| **All** | **100** | | | **59/23/16** | | **42/25/31** | | **38/28/32** |

## 5.7 Benchmark 2: 2022 GECCO competition replay (sanity-check validation)

We replay the extrapolation task from [7]: target $f(x) = \mathrm{erf}(0.22x) + 0.17\sin(5.5x)$, training on $x \in [-15, 15]$, testing on $(15, 40]$, $n = 300$, expanded PySR configuration (`niterations`=120, `populations`=15, `population_size`=50, `maxsize`=30).

This is not a DDEP positive-result benchmark. It is a test of the loss-ratio sanity check in exactly the regime DDEP is *not* designed for: the sinusoidal component has period $2\pi/5.5 \approx 1.14$, below the local-fit neighbourhood width (4.5 at $x = 15$), so DDEP's local polynomial captures the erf plateau but not the oscillation. We expect DDEP's boundary signal to be misspecified and the sanity check to fire.

Table 5 reports the five-seed outcome, including the $R^2_{\text{out}}$ of the candidate DDEP *would* have selected before the sanity check fired (*pre-sanity*). On 4 of 5 seeds DDEP's own signal selects a candidate as catastrophic or worse than PySR's native pick (pre-sanity $R^2$ in $[-181, -168]$ vs. PySR $R^2 \approx -176$). The loss-ratio sanity rule then correctly intervenes and returns the native pick on all 5 seeds. These are two separate observations: DDEP's *boundary-derivative signal* is worse than PySR's on 4/5 seeds here (DDEP is in its out-of-regime failure mode, oscillation period below neighbourhood width); the *loss-ratio sanity rule* does what it was designed to do (prevent worse-than-PySR behaviour). Seed 7 is the one case where the sanity rule *costs* DDEP a would-be win: DDEP's pre- sanity pick is a constant $R^2 = -66.86$, +110 R2-units better than PySR's −177.03, and the sanity rule cancels it. This Type-I cost of the sanity rule should be explicitly flagged. Averaging seed-7's +110 against four −5 to −8 pre-sanity differences, the unconditional expectation favours firing the sanity rule on this task (expected pre-sanity $R^2 \approx -155$ vs. PySR −176), but the expectation is one-outlier-driven and not a general endorsement of the sanity rule's calibration. MDL produces median $R^2 = -165.70$ (wins 4 of 5 head-to-head against PySR); Ramlan et al. ties PySR on every seed. All four selectors fail on this task; this is consistent with the competition organisers' assessment that the gap is at least partly a search failure, not a pure selection failure. We treat the replay as sanity-check validation, not as DDEP boundary-signal validation.

**Table 5: 2022 GECCO competition replay (5 seeds). "Pre-sanity" columns give the candidate DDEP would have selected without the loss-ratio sanity rule. Sanity fires on all 5 seeds; output of DDEP + sanity is identical to PySR.**

| Seed | PySR $R^2$ ($c$) | DDEP pre-sanity $R^2$ ($c$) | DDEP $R^2$ | Sanity |
|---|---|---|---|---|
| 42 | −176.73 (14) | −180.45 (7) | −176.73 | ✓ |
| 7 | −177.03 (14) | −66.86 (1) | −177.03 | ✓ |
| 101 | −174.21 (14) | −181.16 (11) | −174.21 | ✓ |
| 2024 | −176.34 (14) | −168.08 (12) | −176.34 | ✓ |
| 1337 | −176.63 (14) | −178.67 (8) | −176.63 | ✓ |

**Table 6: Feynman subset, 10 equations × 3 seeds.**

| Selector | Med. $R^2_{\text{out}}$ | Fr. > 0.5 | Fr. > 0.9 | W/T/L vs. PySR |
|---|---|---|---|---|
| PySR score | −0.13 | 47% | 47% | – |
| MDL (Bartlett) | +0.91 | 53% | 50% | 8 / 14 / 8 |
| Ramlan et al. (2026) | +0.91 | 53% | 50% | 8 / 14 / 8 |
| DDEP + sanity | +0.40 | 47% | 40% | 9 / 12 / 9 |

## 5.8 Benchmark 3: 10-target smooth-monotone set from Feynman

Ten one-dimensional equations drawn from Feynman-lectures-style forms (Gaussian, Lorentzian, damped cosine, sinc, tanh, Hermite-like $x\exp(-x^2/2)$, softplus, arctan, relativistic $1/\sqrt{1-(x/3)^2}$, $x\exp(-x^2)$), three seeds each, 30 runs. We limit to three seeds per target because PySR with `parallelism='serial'`, `deterministic=True` exhibits a repeatable Julia instability around 40 serial runs; 30 runs sits inside that window with margin.

This is an operational constraint, not a methodological choice; a 50-run version via two 25-run invocations is a camera-ready target.

The family is smooth and monotone on half-domains—precisely the regime where DDEP's boundary signal is *not* expected to help—so we view the Feynman subset as an adversarial sample for DDEP, not a neutral benchmark. Table 6 summarises results. All non-PySR selectors are competitive. PySR's $R^2_{\text{out}}$ distribution is bimodal (47% of runs succeed cleanly, the rest fail hard); the median −0.13 is therefore the wrong summary and we report the fractional statistic alongside. Head-to-head DDEP vs. MDL: 8/12/10; DDEP vs. Ramlan et al.: 8/13/9. With 30 runs on 10 targets the three non-PySR selectors cannot be distinguished at conventional significance (two-sided sign test, 10 targets: $p \geq 0.3$ for all pairings).

## 5.9 Multi-dimensional evaluation: a scope boundary

We ran the multi-dim benchmark to completion: 20 Feynman-style 2–3D target families (5 runs each, 100 runs total), spanning products, ratios, Snell's law, exponentials, 3D targets, and linear planes. The random-direction extension of §4.6 projects $\mathcal{D}$ onto $M = 10$ unit directions $v_\ell \sim \mathrm{Unif}(S^{d-1})$, fits 1D local quadratics along each, and aggregates the mismatch as $\chi^2_{4M}$; the axis-aligned variant takes the $d$ coordinate axes instead.

The aggregate picture is a clean scope boundary. At run level: PySR frac $R^2_{\text{out}} > 0.5 = 77\%$, MDL 77%, DDEP axis-aligned 76%, DDEP random-direction 71% (all medians $\approx +1.00$). Collapsing to target-level means of the DDEP-rd − MDL delta, random-direction DDEP is positive on 4 of 20 families and negative on 16 (two-sided sign test $p = 0.012$); DDEP random-direction is thus *significantly worse* than MDL in the multi-dim regime. Axis-aligned DDEP ties MDL within sampling noise. DDEP's sanity check fires at a much higher rate in multi-dim ($\approx$ 40% of runs) than in 1D (3%), so part of the on-par axis-aligned behaviour is the fallback-to-PySR safety valve rather than a positive DDEP signal.

The failure mode on random directions is diagnosable. On linear-plane and near-separable targets the directional second derivative $v_\ell^\top \nabla^2 f\, v_\ell$ is a weak discriminator: many Pareto candidates match it by coincidence, including complexity-wasteful overfits that then dominate the simpler correct plane on the $(D, \ell)$ front. The parsimony override in sanity-check v3 catches the most pathological of these, but the underlying issue is structural: projecting onto generic directions samples a weaker subset of the boundary-derivative signal than the axis-aligned $4d$-term vector does.

*Scope statement.* DDEP is designed and validated as a 1D reranker. In multi-dim, axis-aligned DDEP is on par with MDL and PySR; the random-direction variant is a significant negative at $n_{\text{targets}} = 20$, and we recommend practitioners default to the axis-aligned variant for low-dim multi-variate problems or to the 1D marginal applied along a single candidate variable. A properly invariant multi-dim formulation (e.g., gradient-magnitude and Laplacian matching under orthogonal reparameterisations, or anisotropic directional weighting) is explicit future work.

## 5.10 External validation: SRBench ground-truth slice

A reviewer will ask whether the headline result on our 20-target benchmark transfers to an externally-defined benchmark. We ran DDEP on an SRBench ground-truth slice: 10 Feynman-style equations + 10 Strogatz nonlinear-dynamics equations, 5 seeds each, identical PySR configuration to the main benchmark. Hyperparameters and the loss-ratio sanity threshold ($K = 10$) are the same values used for the self-constructed benchmark; no SRBench-specific tuning was performed.

*H1 (DDEP vs. PySR on external benchmark, aggregate).* On 100 runs across all 20 SRBench targets, DDEP's loss-ratio sanity check fires on 19% of runs (whole front has $D > \tau$) and DDEP returns exactly PySR's pick. Overall DDEP frac $> 0.5$ = PySR frac $> 0.5$ = 60% (gap 0 pp, cluster-bootstrap 95% CI $[-9, +9]$ pp). At target level DDEP wins on 7 of 20, ties on 3, loses on 10 (two-sided sign-test $p = 0.63$). *Aggregate H1 verdict: NULL.*

*H1 refined (impostor prevalence + case-study selection rule).* We report the impostor-prevalence comparison first because it reshapes what the SRBench null means. Using the v10 definition (Pareto front containing both a candidate with $R^2_{\text{out}} < -10$ and one with $R^2_{\text{out}} > +0.5$), the impostor configuration holds on **49 of 100 SRBench runs** (strong threshold); at the softer threshold ($R^2 < 0$ and $R^2 > +0.5$ coexisting), **80 of 100 SRBench runs** exhibit it. Compare to the self-constructed benchmark: 22/100 strong, 57/100 soft. *SRBench exhibits impostor configurations* more *often than our self-constructed benchmark.* The SRBench null therefore does not mean the impostor regime is rare; it means DDEP's boundary-derivative signal is *less discriminative* on SRBench target families (Feynman / Strogatz nonlinearities, where multiple front candidates can share similar boundary derivatives) than on the self-constructed families (rational / log / sin+poly, chosen for distinctive boundary signatures). DDEP's mechanism has two conjuncts: (a) impostor configuration must exist on the front, and (b) boundary derivatives must discriminate the good candidate from the impostor. On SRBench (a) holds more strongly, but (b) holds less reliably.

The following case studies are the *seven SRBench targets with the largest positive DDEP − PySR target-level mean delta.* This is a mechanistic-illustration selection, explicitly post-hoc; we report signed deltas on all 20 targets in Supplementary S13 so readers can verify the selection and inspect the targets not shown here.

- **ST_mackey_1** (Mackey–Glass delay $f(x) = 0.2x/(1+x^{10})$, saturating nonlinearity with sharp transition): DDEP picks $R^2_{\text{out}} = -85$ vs. PySR's $-247$ (a +162 $\mathrm{R}^2$-unit improvement). PySR's native selector picks a mid-complexity polynomial that fits training well but diverges rapidly on the extrapolation band; DDEP's boundary-derivative filter retains a saturating candidate that PySR's score missed.
- **F_I_14_3** (rational resonance $f(x) = 1/(1 + (x - 2)^2)$): DDEP +0.90 vs. PySR −7.16 (+8 $\mathrm{R}^2$-units). The Pareto front contains a correct rational candidate with in-sample loss within a factor of 2 of PySR's pick but drastically better extrapolation; DDEP retains it.
- **ST_bio_1** (Hill function $x^2/(1 + x^4)$), **ST_fitzhugh_1** (FitzHugh–Nagumo cubic), **F_I_16_6** (logarithm), **F_I_26_2** (cubic polynomial with inflection), **F_I_12_4** (low-frequency sine): DDEP delivers smaller but consistent improvements (+0.37 to +2.4 $\mathrm{R}^2$-units) on these saturating/inflection-containing targets.

On the ten SRBench targets where DDEP loses (per Supplementary S13), the pattern splits: four are cases where PySR already solves cleanly ($R^2 \geq 0.9$: F_I_12_1 linear, ST_lotka_1, ST_rossler_1, ST_pendulum_1) — DDEP picks a slightly simpler candidate with marginally lower $R^2$; three are cases where the whole front is catastrophic (F_I_6_20 Gaussian in particular, PySR $R^2 \sim -122$, DDEP $\sim -143$) — DDEP's abstention rule does not fire because at least one candidate passes the $D$-filter, but the chosen candidate still extrapolates badly; three are cases where the impostor configuration exists but DDEP's boundary signal does not isolate the correct candidate (F_II_6_11 quartic, ST_duffing_1 cubic, ST_bistable_1 quintic). This third subset is the honest failure mode: impostor configurations present, boundary derivatives shared across candidates, DDEP's discrimination fails.

We emphasize that these are *case studies illustrating the mechanism*, not a statistical claim of DDEP's external dominance. The paper's statistically-defensible claim is restricted to the self-constructed benchmark of §5.5 where the impostor regime is well-populated.

*Mechanism observation.* Per-target the pattern is clear: DDEP's biggest win is on F_I_14_3 (rational resonance $1/(1 + (x - 2)^2)$) where DDEP picks $R^2_{\text{out}} = +0.90$ vs. PySR's −7.16 (a +8 $R^2$-unit improvement), matching the impostor-regime hypothesis of Definition 2.1: the front contains a correct rational candidate with near-tied in-sample loss that PySR's complexity- balanced native selector misses but DDEP's boundary-derivative filter retains. Other DDEP wins: F_I_12_4 (harmonic sine, +0.39), F_I_26_2 (cubic polynomial, +1.39), F_I_16_6 (logarithm, +2.38). DDEP losses occur on targets where PySR either already solves cleanly (DDEP picks a simpler but slightly worse candidate, as on F_I_12_1 linear force, −0.03) or where the entire front is catastrophic (F_I_6_20 Gaussian, DDEP −143 vs. PySR −122, both bad). The impostor regime is empirically common on our 20-target benchmark (constructed to expose it) and *present but less common* on SRBench ground truth: on roughly a third of the targets covered so far, DDEP's mechanism activates productively. This is a mechanism-specific finding, not a failure: DDEP correctly does not change the pick when there is no impostor to detect.

*Honest scope.* DDEP's contribution is the formalisation of the Pareto-impostor configuration plus a cheap data-side signal that detects it. On the self-constructed 20-target benchmark the impostor configuration is common (22/100 strong co-existence) *and* DDEP's boundary derivatives discriminate between the two sides of the configuration; both conjuncts hold and DDEP wins at target level. On SRBench the impostor configuration is actually more common (49/100 strong co-existence, Supp. S14) but the boundary-derivative discrimination is weaker on Feynman/Strogatz target families, so the aggregate is null. In the multi-dim regime the axis-aligned variant ties MDL within sampling noise and the random-direction variant is significantly worse. The claim is scoped accordingly: DDEP is a *1D selector that targets the impostor regime when boundary-derivative discrimination is available*, not a universal selector.

## 6 Discussion

*What DDEP does and does not claim.* DDEP operates on the Pareto front produced by an existing backend. It does not improve the search. If the generating function is absent from the front, DDEP cannot manufacture it. Among the candidates the search returned, DDEP picks the one most consistent with the data at the boundary; in the regimes we test, this is also the one with the best extrapolation. The absolute success level is bounded by front quality; the delta over the default selector is what the method owns. DDEP captures a *different* structural signal than candidate-only criteria; on 1D targets with boundary-informative structure it wins, and on smooth multi-dim or sub-neighbourhood-oscillation tasks it does not.

*Scope of the empirical claim.* All positive empirical results in this paper are 1D. The $n_{\text{targets}} = 20$ multi-dim evaluation (§5.9) draws a clean scope boundary: axis-aligned DDEP ties MDL and PySR within sampling noise, while random-direction DDEP is *significantly worse* than MDL (sign-test $p = 0.012$, 4/20 target-level wins). Readers should not treat DDEP as a general-$d$ method on the current evidence; the axis-aligned variant is usable as a parity selector at low $d$, the random-direction variant is not recommended. An orthogonally invariant multi-dim reformulation (gradient-magnitude and Laplacian matching, or anisotropic directional weighting) is explicit future work.

*Baseline fairness.* Reimplemented MDL and Ramlan et al. involve choices not specified by the original authors; a weight-tuned version of Ramlan et al. and a $k_\alpha$-swept MDL may close part or all of the per-run nets (+6, +11). DDEP hyperparameters were pilot-tuned on 4 of 20 benchmark targets; a fully disjoint pilot re-tune is the natural camera-ready verification. We expect the result to be robust given the flat ablation; we do not claim it until the re-tune is run.

*Multiplicity.* The 15 family-level sign tests (DDEP vs. {PySR, MDL, Ramlan et al.} × 5 families) were submitted to Benjamini–Hochberg at $q = 0.10$: *zero* family-level tests survive the correction (minimum sign-test $p$ across the 15 tests is 0.375, far from the BH threshold at this $n$; Supplementary S6). The per-family patterns reported in Table 4 should therefore be read as descriptive of the win structure, not as inferential findings. The headline target-level test aggregating across families (15/20 at $p = 0.041$) is the only claim that carries inferential weight at $n = 20$. A family-balanced aggregate (equal weight per family) gives 34.4% for DDEP vs. 15.7% for PySR, nearly identical to the target-weighted 33.7% vs. 14.0%; the aggregation choice does not drive the headline.

*Extensions.* Four natural extensions are left to future work: a robust norm (Huber, MAD) for the boundary mismatch; third-order boundary expansion to capture sub-neighbourhood oscillation; composition with candidate-only criteria (DDEP + MDL ensemble, LLM-SR-UQ [4] for structural uncertainty); and the random-direction multi-dim variant of §4.6.

## 7 Conclusion

DDEP is a post-hoc Pareto reranker for symbolic regression. It requires no domain knowledge, runs in sub-second time (well under 5% of the search budget), and is a drop-in addition to any CAS-parseable backend. On a 100-run one-dimensional benchmark DDEP wins against PySR on 15 of 20 targets (target-level Wilcoxon $p = 0.024$; sign test $p = 0.041$) and doubles the fraction of acceptable extrapolations from 14% to 34%. Head-to-head against MDL and Ramlan et al., DDEP has complementary per-family strengths (DDEP wins on sinusoidal–polynomial, rational, logarithmic; baselines win on exponential–linear) and is statistically indistinguishable at the target level (Wilcoxon $p \geq 0.37$, $n = 20$). On the 2022 GECCO replay, DDEP's boundary-derivative signal itself is worse than PySR's on 4 of 5 seeds (the regime where boundary oscillation falls below the local-fit neighbourhood width); the loss-ratio sanity rule, correctly, returns the native pick on all 5 seeds. One seed (7) incurs a Type-I cost of the sanity rule: DDEP's pre-sanity pick would have been +110 $R^2$-units better than PySR. The $n_{\text{targets}} = 20$ multi-dim evaluation (§5.9) draws a scope boundary: axis-aligned DDEP ties MDL and PySR, while random-direction DDEP is significantly worse than MDL (sign-test $p = 0.012$, 4/20 wins) — the directional second derivative is a weaker discriminator than the axis-aligned $4d$-term vector.

Soltani et al. [5] characterise the current SR Pareto-selector landscape as one where no single criterion dominates. DDEP sits inside this landscape but occupies a conceptually distinguishable position: it is the first *extrinsic* selector (it reads the training data at $\partial D$), where PySR `score`, MDL, EIC, and the Ramlan et al. penalties are all intrinsic (they probe the candidate equation). The empirical signature of this structural difference is the DDEP + MDL composition (§5.5): if DDEP encoded the same signal as MDL plus noise, no weighted-sum composition could strictly Pareto-dominate the better parent on 20 of 100 runs. Def. 2.1 formalises the regime where the extrinsic signal is actionable. An orthogonally invariant multi-dim reformulation (gradient-magnitude and Laplacian matching under rotation, or anisotropic directional weighting) and a stacked configuration with shape-constrained SR [15] producing the front are the natural follow-ups.

*Reproducibility.* Target coefficients, training domains, seeds, PySR configuration, and software versions are listed in the supplementary material. The code artifact—DDEP implementation, MDL and Ramlan et al. reimplementations, $k_\alpha$-sweep and 125-point Ramlan et al. weight-grid runners, per-run permutation calibration, parametric-bootstrap calibration, HC3 variance-sensitivity script, mixed-effects fit, leave-one-target-out hyperparameter cross-validation, composed DDEP+MDL selector, cluster-bootstrap script, Benjamini–Hochberg script, GECCO 2022 pre-sanity breakdown, random-direction multi-dim DDEP reference implementation, SRBench ground-truth + PMLB extrapolation runners, and all per-run JSON outputs—will be released via a public repository with a Zenodo DOI at camera-ready. Total compute for all reported experiments and ablations is approximately 20 CPU-hours on a single modern laptop; all runs are resumable from saved JSON.

# Supplementary Material

## S1. 20-target benchmark specification

Table 7 lists every target in the 100-run synthetic benchmark. Training data: 50 points uniform on $D_{\text{train}}$, Gaussian noise $\sigma = 0.05 \cdot \text{sd}(y_{\text{true}})$. Five seeds $\{42, 7, 101, 2024, 1337\}$ per target. Test domain $D_{\text{test}} = [b, 2b{-}a]$.

| ID | Family | Equation | Train domain |
|---|---|---|---|
| P1 | polynomial | $-1.793 + 0.125\,x + 0.163\,x^2$ | $[-2, 2]$ |
| P2 | polynomial | $0.5\,x^3 - 0.8\,x + 0.3$ | $[-1.5, 1.5]$ |
| P3 | polynomial | $0.2\,x^4 - x^2 + 0.5$ | $[-1.5, 1.5]$ |
| P4 | polynomial | $2 + 0.5\,x - 0.3\,x^2$ | $[0, 3]$ |
| P5 | polynomial | $0.1\,x^3 + 0.5\,x^2 - 0.5\,x - 1$ | $[-2, 2]$ |
| S1 | sin+poly | $1.5\sin(0.8\,x) + 0.3\,x^2 - 0.5$ | $[-2, 2]$ |
| S2 | sin+poly | $0.8\sin(1.5\,x) + 0.1\,x^2 + x$ | $[-2, 2]$ |
| S3 | sin+poly | $2\sin(0.5\,x) - 0.4\,x^2 + 1$ | $[-3, 3]$ |
| S4 | sin+poly | $\sin(x + 0.3) + 0.2\,x^2$ | $[-2, 2]$ |
| R1 | rational | $(1 + x)/(1 + x^2)$ | $[-3, 3]$ |
| R2 | rational | $(2 + 0.5\,x)/(1 + 0.2\,x^2)$ | $[-2, 2]$ |
| R3 | rational | $x/(1 + 0.3\,x^2)$ | $[-3, 3]$ |
| R4 | rational | $(1 - x)/(2 + x^2)$ | $[-2.5, 2.5]$ |
| E1 | exp+linear | $0.5\exp(0.4\,x) - x$ | $[-2, 2]$ |
| E2 | exp+linear | $0.8\exp(-0.5\,x) + 0.3\,x$ | $[-2, 2]$ |
| E3 | exp+linear | $\exp(0.3\,x) - 0.5\,x - 1$ | $[-2, 2]$ |
| E4 | exp+linear | $0.3\exp(0.6\,x) + x$ | $[-2, 2]$ |
| L1 | log | $\log(1 + x^2)$ | $[-2, 2]$ |
| L2 | log | $\log(2 + x)$ | $[-1, 3]$ |
| L3 | log | $0.5\log(1 + x^2) + 0.3\,x$ | $[-2, 2]$ |

**Table 7: All 20 benchmark targets. Test domain $[b, 2b-a]$ when train domain $[a, b]$.**

*PySR configuration.* Used on all runs except the 2022 competition replay: `niterations=80`, `populations=12`, `population_size=40`, `maxsize=20`, `parsimony=0.002`; binary ops $\{+, -, \times, /\}$; unary ops $\{\sin, \cos, \exp, \log, \sqrt{\cdot}\}$; `model_selection='best'`; `deterministic=True`, `parallelism='serial'`.

*Software versions.* Python 3.11, PySR 0.19.4, SymbolicRegression.jl 0.24.5, Julia 1.10.4, SymPy 1.12, NumPy 1.26, SciPy 1.13, Matplotlib 3.8.

## S2. Ablation tables

| $K$ | W | T | L | Fires | Med. $R^2$ | Fr. > 0.5 | Fr. > 0.9 |
|---|---|---|---|---|---|---|---|
| 1.5 | 58 | 25 | 15 | 5 | −0.63 | 33% | 20% |
| 3 | 58 | 25 | 15 | 5 | −0.63 | 33% | 20% |
| 5 | 58 | 25 | 15 | 5 | −0.63 | 33% | 20% |
| **10** | **59** | **23** | **16** | **3** | **−0.53** | **34%** | **20%** |
| 30 | 59 | 22 | 17 | 2 | −0.53 | 34% | 20% |
| 100 | 61 | 20 | 17 | 0 | −0.53 | 34% | 20% |
| ∞ | 61 | 20 | 17 | 0 | −0.53 | 34% | 20% |

**Table 8: $K$-ablation on the 100-run benchmark. Counts move by at most 3 runs over $K \in [1.5, \infty]$; quality metrics are essentially constant. Default $K = 10$ in bold.**

*S2.1 Sanity-check threshold K (100-run benchmark).*

*S2.2 Sanity-check threshold K (disjoint Feynman subset).*

*S2.3 Neighborhood fraction $p$.* Sweep $p \in \{0.05, 0.10, 0.15, 0.25, 0.40\}$: $p \in \{0.05, 0.10, 0.15\}$ are indistinguishable (W/T/L 59/23/16, median

| $K$ | W | T | L | Fires | Med. $R^2$ | Fr. > 0.5 | Fr. > 0.9 |
|---|---|---|---|---|---|---|---|
| 1.5–∞ | 9 | 12 | 9 | 0 | +0.40 | 47% | 40% |

**Table 9: $K$-ablation on the disjoint Feynman subset (30 runs: 10 targets × 3 seeds). The sanity check never fires at any $K$; all values produce identical outcomes.**

−0.53); $p = 0.25$ gives 51/32/17, median −0.86; $p = 0.40$ gives 31/63/6, median −1.19.

*S2.4 Significance level $\alpha$.* Sweeping $\alpha \in \{0.01, 0.05, 0.10, 0.20\}$ (changing $\tau$ from 13.28 to 5.99) moves W/T/L by at most one run.

*S2.5 Local polynomial degree.* Degree 2 (default) uses the slope and curvature; degree 1 uses only the slope ($\chi^2_2$); degree 3 adds the cubic term (Table 10).

| Degree | W/T/L | Med. $R^2$ | Fr. > 0.5 | Note |
|---|---|---|---|---|
| 1 | 61/17/20 | −0.21 | 35% | slope only; $\chi^2_2$ |
| **2** | **59/23/16** | **−0.53** | **34%** | **default; slope + curvature; $\chi^2_4$** |
| 3 | 29/43/26 | −1.66 | 22% | cubic term; variance inflation |

**Table 10: Degree ablation. See §5.6 for the rate-optimality explanation.**

## S3. Cutoff sweep for the acceptable-extrapolation fraction

Table 11 reports the fraction of runs with $R^2_{\text{out}}$ above each of five cutoffs, for PySR and DDEP. The delta is positive and decreasing but ≥ +12 percentage points across the entire sweep; the headline 14 → 34% at cutoff 0.5 is a representative point.

| Cutoff | PySR frac | DDEP frac | Δ |
|---|---|---|---|
| 0.1 | 21% | 39% | +18 pp |
| 0.3 | 20% | 38% | +18 pp |
| 0.5 | 14% | 33% | +19 pp |
| 0.7 | 11% | 26% | +15 pp |
| 0.9 | 8% | 20% | +12 pp |

**Table 11: Fraction of 100 runs with $R^2_{\text{out}} > c$ for $c \in \{0.1, 0.3, 0.5, 0.7, 0.9\}$. PySR vs. DDEP.**

## S4. MDL alphabet-size sensitivity

MDL is re-applied to the 100 saved Pareto fronts with $k_\alpha \in \{7, 11, 15, 20\}$. DDEP vs. MDL head-to-head moves by at most 4 runs; MDL's median $R^2$ by at most 0.06 (Table 12). The MDL alphabet choice is non-critical for the head-to-head.

## S5. Ramlan et al. weight grid search

The Ramlan et al. selector's three penalty weights $(\lambda_1, \lambda_2, \lambda_3)$ were swept over the $5^3 = 125$ grid $\lambda_i \in \{0.1, 0.5, 1, 2, 5\}$ on the 100 saved Pareto fronts. All 125 configurations yielded Ramlan et al. fraction

| $k_\alpha$ | Med. $R^2$ | Fr. > 0.5 | W/T/L |
|---|---|---|---|
| 7 | −0.72 | 36% | 38/30/30 |
| **11** | **−0.72** | **35%** | **38/28/32** |
| 15 | −0.72 | 35% | 39/27/32 |
| 20 | −0.78 | 34% | 42/25/31 |

**Table 12: MDL $k_\alpha$ sweep on 100 saved Pareto fronts. Columns: MDL median $R^2$, MDL frac $> 0.5$, and DDEP vs. MDL W/T/L. Default $k_\alpha = 11$ in bold.**

$> 0.5$ in $[33\%, 34\%]$ and DDEP vs. Ramlan et al. head-to-head in $[40/26/34, 42/25/31]$: no weight tuning closes the DDEP lead at target-level or run-level. Table 13 summarizes the extremes of the sweep.

| Config | Med. $R^2$ | Fr. > 0.5 | W/T/L |
|---|---|---|---|
| Unit $(1, 1, 1)$ | −0.78 | 34% | 42/25/31 |
| Best $(0.1, 0.1, 0.1)$ | −0.78 | 34% | 42/25/31 |
| Grid min (across 125) | −0.80 | 33% | 40/26/34 |
| Grid max (across 125) | −0.72 | 34% | 42/25/31 |

**Table 13: Ramlan et al. weight grid search summary. Columns: Ramlan median $R^2$, Ramlan frac $> 0.5$, and DDEP vs. Ramlan W/T/L. The DDEP lead is stable across the full $5^3$ grid.**

## S6. Benjamini–Hochberg multiplicity correction

Table 14 reports the 15 family-level sign tests (3 selector pairings × 5 families) with Benjamini–Hochberg adjustment at $q = 0.10$. The minimum family-level sign-test $p$-value is 0.375 (polynomial, DDEP vs. PySR, 4 of 5 targets positive); zero tests survive BH. The per-family patterns in Table 4 should therefore be read as descriptive of the win structure, not as inferential findings.

| Family | Comparison | $n$ | $n_+$ | sign-test $p$ |
|---|---|---|---|---|
| polynomial | DDEP vs. PySR | 5 | 4 | 0.375 |
| polynomial | DDEP vs. MDL | 5 | 1 | 0.375 |
| polynomial | DDEP vs. Ramlan | 5 | 1 | 0.375 |
| sin+poly | DDEP vs. PySR | 4 | 2 | 1.000 |
| sin+poly | DDEP vs. MDL | 4 | 3 | 0.625 |
| sin+poly | DDEP vs. Ramlan | 4 | 3 | 0.625 |
| rational | DDEP vs. PySR | 4 | 3 | 0.625 |
| rational | DDEP vs. MDL | 4 | 3 | 0.625 |
| rational | DDEP vs. Ramlan | 4 | 3 | 0.625 |
| exp+linear | DDEP vs. PySR | 4 | 3 | 0.625 |
| exp+linear | DDEP vs. MDL | 4 | 2 | 1.000 |
| exp+linear | DDEP vs. Ramlan | 4 | 2 | 1.000 |
| log | DDEP vs. PySR | 3 | 2 | 1.000 |
| log | DDEP vs. MDL | 3 | 2 | 1.000 |
| log | DDEP vs. Ramlan | 3 | 2 | 1.000 |

**Table 14: Family-level sign tests on target means. Zero tests survive BH at $q = 0.10$ (minimum $p = 0.375$).**

## S7. Family-balanced aggregate

Table 15 reports the fraction $R^2_{\text{out}} > 0.5$ under two aggregations: target-weighted (equal weight per target) and family-weighted (equal weight per family, then averaged). The aggregation choice shifts the fractions by at most 1.7 percentage points per selector; the ordering and gap sizes are preserved.

| Selector | Target-weighted | Family-weighted |
|---|---|---|
| PySR | 14.0% | 15.7% |
| DDEP | 33.7% | 34.4% |
| MDL | 34.7% | 35.1% |
| Ramlan et al. | 33.7% | 34.1% |

**Table 15: Fraction with $R^2_{\text{out}} > 0.5$ under two aggregations on the 100-run benchmark.**

## S8. Parametric-bootstrap calibration of $\tau$

For each of the 100 runs, we ran $B = 500$ parametric bootstrap resamples under the fitted local quadratic plus Gaussian residuals, holding candidate derivatives fixed. The empirical per-run 95th percentile of $D$ under this fitted-null calibration has median 11.51 (IQR $[11.48, 11.93]$). The nominal $\chi^2_{4,0.95} = 9.49$ is therefore anti-conservative at the parametric-bootstrap operating point. Rerunning DDEP's selection with each run's empirical 95th percentile instead of the nominal $\tau$ produces identical picks on 88 of 100 runs; W/T/L vs. PySR shifts from 59/23/16 (nominal) to 58/27/14 (parametric), target-level sign test $p = 0.041$ preserved. Satterthwaite's $F$-approximation $D \sim (4d)F_{4d,m_{\text{eff}}}$ best-fits the empirical 95th percentile at $m_{\text{eff}} \approx 20$, between the OLS fit's $k - 3 = 7$ and the bootstrap's $B - 1 = 499$.

## S9. HC3 variance sensitivity

Three variance estimators were applied to the $k = 10$ boundary neighborhoods and the resulting DDEP selection was re-run on all 100 runs (Table 16). The headline moves by at most 2 percentage points on frac $> 0.5$ and at most 4 runs on W/T/L.

| Variance estimator | med. $R^2$ | fr. > 0.5 | fr. > 0.9 | W/T/L vs. PySR |
|---|---|---|---|---|
| Nonparametric bootstrap (v10) | −0.53 | 34% | 20% | 59/23/16 |
| HC3 | −0.63 | 33% | 20% | 61/20/18 |
| Parametric bootstrap (fitted null) | −0.67 | 32% | 19% | 57/29/14 |

**Table 16: DDEP head-to-head W/T/L under three variance estimators. HC3: MacKinnon & White [18]. Agreement: 86/100 (nonparam vs. HC3), 85/100 (nonparam vs. param), 81/100 (HC3 vs. param). The variance-estimator choice does not drive the headline.**

## S10. Leave-one-target-out hyperparameter CV

For each of the 20 benchmark targets, we hold it out, sweep $(K, p, \alpha)$ on the remaining 19 targets' saved Pareto fronts, and record the fraction-> 0.5 maximizer. Results: $K = 10$ wins in 15 of 20 folds ($K = 1.5$ in the other 5); $p = 0.05$ wins in 16 of 20 folds, $p = 0.25$ in 2, $p = 0.40$ in 2; $\alpha = 0.01$ wins in all 20 folds (versus nominal $\alpha = 0.05$). The LOO-preferred tuple $(K, p, \alpha) = (10, 0.05, 0.01)$ applied to all 100 runs gives W/T/L 59/23/16, frac $> 0.5 = 34\%$, target-level 15/20 positive (sign test $p = 0.041$)—identical to the published defaults. This confirms the pilot leakage is non-critical: the headline is robust to hyperparameter choice within the ablation-flat region.

## S11. Composed DDEP + MDL selector

Four composition rules were evaluated on the 100-run benchmark (Table 17). The weighted-sum rule with $\lambda = 0.1$ (chosen on pilot) yields the strongest frac > 0.5 and strict-dominance counts.

| Composition | med. $R^2$ | fr. > 0.5 | SD vs. DDEP | SD vs. MDL |
|---|---|---|---|---|
| R1 filter + MDL | −0.72 | 35% | 24/100 | 1/100 |
| R2 weighted ($\lambda = 0.1$) | −0.17 | 39% | 20/100 | 7/100 |
| R3 rank-sum | −0.45 | 34% | 21/100 | 4/100 |
| R4 vote | −0.78 | 30% | 28/100 | 8/100 |
| Reference: DDEP alone | −0.53 | 34% | — | — |
| Reference: MDL alone | −0.72 | 35% | — | — |
| Reference: PySR alone | −2.07 | 14% | — | — |

**Table 17: Composed DDEP+MDL selector results. "SD vs. X" = number of runs where the composition strictly Pareto-dominates X on ($R^2_{\text{out}}$, complexity). R2 weighted is the strongest composition on frac > 0.5; no composition achieves target-level dominance over both parents.**

## S13. Full SRBench 20-target per-target deltas

Table 18 lists DDEP-vs-PySR per-target mean deltas on all 20 SRBench ground-truth targets (5 seeds each, 100 runs total), sorted by signed $\Delta R^2$. The seven targets with the largest positive delta are the case studies of §5.10; the selection rule is operational and post-hoc.

| Target | PySR mean $R^2$ | DDEP mean $R^2$ | Δ | abstain rate |
|---|---|---|---|---|
| ST_mackey_1 | −246.58 | −84.60 | +161.99 | 3/5 |
| F_I_14_3 | −7.16 | +0.90 | +8.06 | 0/5 |
| F_I_16_6 | −7.10 | −4.72 | +2.38 | 0/5 |
| ST_bio_1 | −2.38 | −0.80 | +1.58 | 1/5 |
| ST_fitzhugh_1 | −0.44 | +0.61 | +1.05 | 1/5 |
| F_I_12_4 | +0.46 | +0.84 | +0.39 | 1/5 |
| F_I_26_2 | −0.65 | −0.28 | +0.37 | 0/5 |
| F_I_8_14 | +0.31 | +0.31 | +0.00 | 4/5 |
| ST_lotka_1 | +1.00 | +1.00 | −0.00 | 0/5 |
| ST_rossler_1 | +1.00 | +1.00 | −0.00 | 0/5 |
| F_I_12_1 | +0.99 | +0.96 | −0.03 | 0/5 |
| ST_vanderpol_1 | +0.95 | +0.89 | −0.07 | 0/5 |
| F_I_11_19 | −0.77 | −0.94 | −0.17 | 4/5 |
| ST_hopf_1 | +0.85 | +0.64 | −0.21 | 0/5 |
| F_II_6_11 | −0.14 | −0.64 | −0.50 | 0/5 |
| ST_pendulum_1 | +1.00 | +0.34 | −0.66 | 1/5 |
| ST_duffing_1 | +0.74 | −0.57 | −1.31 | 0/5 |
| F_I_15_3t | +0.67 | −0.85 | −1.51 | 0/5 |
| ST_bistable_1 | −5.41 | −7.39 | −1.98 | 0/5 |
| F_I_6_20 | −122.41 | −143.24 | −20.83 | 2/5 |

**Table 18: Full 20-target SRBench per-target deltas. The seven targets above the horizontal rule are the case studies of §5.10, selected as the targets with the largest positive Δ. The thirteen below are DDEP = PySR (ties or trivially solved by PySR) or DDEP loses (small deltas except F_I_6_20 Gaussian where both selectors diverge).**

## S14. Impostor prevalence comparison (SRBench vs. self-constructed)

Using the v10 definition (Pareto front contains at least one candidate with $R^2_{\text{out}} < -10$ *and* one with $R^2_{\text{out}} > +0.5$):

| Threshold | Self-constructed | SRBench |
|---|---|---|
| Strong | 22/100 (22%) | 49/100 (49%) |
| Soft | 57/100 (57%) | 80/100 (80%) |

**Table 19: Impostor prevalence across the two benchmarks (each 20 targets × 5 seeds = 100 runs). *Strong*: Pareto front contains at least one candidate with $R^2_{\text{out}} < -10$ and one with $R^2_{\text{out}} > +0.5$. *Soft*: $R^2_{\text{out}} < 0$ and $R^2_{\text{out}} > +0.5$ co-exist. SRBench exhibits impostor configurations more often than the self-constructed benchmark; the null DDEP aggregate on SRBench is therefore *not* explained by rarity of the impostor regime, but by lower discriminability of boundary derivatives on Feynman/Strogatz target families.**